\documentclass[prl,a4paper,twocolumn,superscriptaddress,longbibliography]{revtex4-2}

\usepackage{amssymb}
\usepackage{amsmath}
\usepackage{amsfonts}
\usepackage{graphicx}
\usepackage{bm}
\usepackage[usenames,dvipsnames]{xcolor}
\usepackage{multirow}
\usepackage{comment}
\usepackage{pgfplots}
\usepackage{tikz}

\usepackage[colorlinks,citecolor=blue,linkcolor=red,urlcolor=blue]
{hyperref}

\usepackage[final]{pdfpages}
\usepackage{pgffor}

\newcommand{\sqH}{{sqH}}

\newcommand{\iqH}{{iqH}}
\newcommand{\NLSM}{{NL$\sigma$M}}

\newcommand{\ECs}{{ECs}}

\makeatletter
\AtBeginDocument{\let\LS@rot\@undefined}
\makeatother

\makeatletter
\newcommand\footnoteref[1]{\protected@xdef\@thefnmark{\ref{#1}}\@footnotemark}
\makeatother

\begin{document}

\title{Emergent spin quantum Hall edge states at the boundary of two-dimensional electron gas proximitized by  an $s$-wave superconductor}

\author{M. V. Parfenov}

\affiliation{\mbox{L. D. Landau Institute for Theoretical Physics, Semenova 1-a, 142432, Chernogolovka, Russia}}

\affiliation{Laboratory for Condensed Matter Physics, HSE University, 101000 Moscow, Russia}

\affiliation{P. N. Lebedev Physical Institute, Russian Academy of Sciences, 119991 Moscow,
Russia}

\author{V. S. Khrapai}

\affiliation{Osipyan Institute of Solid State Physics, Russian Academy of Sciences, 142432 Chernogolovka, Russia}

\affiliation{HSE University, 101000 Moscow, Russia}

\author{I. S. Burmistrov}

\affiliation{\mbox{L. D. Landau Institute for Theoretical Physics, Semenova 1-a, 142432, Chernogolovka, Russia}}

\affiliation{Laboratory for Condensed Matter Physics, HSE University, 101000 Moscow, Russia}

\affiliation{P. N. Lebedev Physical Institute, Russian Academy of Sciences, 119991 Moscow, Russia}

\begin{abstract}
Hybrid two-dimensional electron gas -- superconductor (2DEG-S) structures in a quantized magnetic field offer a promising platform for realizing new topological phases. While recent experiments reveal chiral Andreev edge states, their charge conductance is not integer quantized and is disorder sensitive, raising the question of whether topological protection survives. We argue that it does, but manifests in the spin transport channel. The 2DEG-S system belongs to symmetry class C of the Altland-Zirnbauer classification, which supports an even-integer quantized transverse spin conductivity -- the spin quantum Hall effect, so far unobserved experimentally. We demonstrate that 2DEG-S hybrids host topologically protected edge states carrying a spin current with an even-integer quantized spin conductance robust against disorder. Finally, we propose an experimental setup to probe this protection via electrical measurements, establishing a concrete route to detect the class C origin of the chiral Andreev edge states.
\end{abstract}

\date{version 6, \today }

\maketitle


The realization of new topological phases beyond established paradigms remains a major challenge in condensed matter physics. Topological states of matter support protected edge modes that enable dissipationless transport as exemplified by the integer quantum Hall ({\iqH}) effect in two-dimensional electron gas (2DEG) \cite{Klitzing1980,Tsui1981} and the quantum spin Hall effect in two-dimensional topological insulators \cite{Kane2005,Bernevig2006,Molenkamp2007}. In these systems, the structure of edge states is dictated by bulk topology. A promising route to new phases is the controlled interfacing of distinct topological systems. Such hybridization gives rise to the topological proximity effect \cite{Hsieh2016,Cheng2019,Panas2020}, in which coupling reshapes both bulk invariants and boundary excitations. Engineered interfaces can thus host artificial edge states, enabling the design of topological phases from well-understood building blocks \cite{Ronen2018,Gefen2019}.

An intriguing fundamental question is the interplay of the ordinary and topological proximity 
effects. A natural setting to explore this interplay is a superconductor (S) proximity-coupled to 2DEG in a perpendicular magnetic field. Although superconductor--normal interfaces in magnetic fields have been extensively studied in the past \cite{hoppe2000,Zulicke2005,hoppe2001,giazotto2005,Chtchelkatchev2007,khaymovich2010}, their potential for hosting exotic topological states with Majorana and parafermionic zero modes has only recently been demonstrated \cite{Mong2014,Clarke2014}.  

Recent experiments \cite{Lee2017,Zhao2020,Shabani2022,Zhao2023,Zhao2024} have shown that a superconductor modifies the structure of {\iqH} edge states through electron-hole conversion due to Andreev reflections at a normal metal-superconductor interface. The resulting chiral Andreev edge states \cite{Zhao2020} do not demonstrate integer quantized charge conductance in units of $G_0=2e^2/h$, where $e$ is electron charge and $h$ is Planck's constant, unlike their parent {\iqH} edge states. 
Furthermore, disorder in the superconductor induces strong fluctuations in the Andreev reflection probability and, consequently, in the conductance of a chiral Andreev edge state~\cite{glazman2023}. These observations
raise a critical question: \emph{does the topological protection of the {\iqH} edge states disappear, as the non-quantized charge conductance suggests, or does it persist in 2DEG-S hybrid structure?}

In this Letter, we argue that the answer to this question is affirmative. The key observation is that a superconductor proximity-coupled to 2DEG in a quantized magnetic field falls into symmetry class C of the Altland--Zirnbauer classification \cite{Zirnbauer1997}. A 2DEG with class C symmetry is known to exhibit an even-integer quantized transverse spin conductivity: the so-called spin quantum Hall ({\sqH}) effect \cite{Volovik1997, Gruzberg1999, Senthil1999, Read2000, Kagolovsky1999}, which is governed by an integer-valued bulk topological invariant \cite{Kitaev2009,Schnyder2009}. However, unlike the {\iqH} effect, the {\sqH} effect has so far eluded experimental observation, underscoring the need for minimal and controllable realizations of class C edge states. The $2\mathbb{Z}$ topological invariant of class C ensures the robustness of edge states in 2DEG-S hybrid structures. Below, we  
 show that  chiral Andreev edge states in such hybrids  
are nothing but {\sqH} edge modes carrying spin current with an even-integer quantized spin conductance that remains robust against disorder. We propose an experimental setup in which the topological protection of these {\sqH} edge states can be probed via electrical measurements.

\begin{figure}[t!]
\begin{center}
\vspace{0mm}
 \includegraphics[width=0.9\columnwidth]{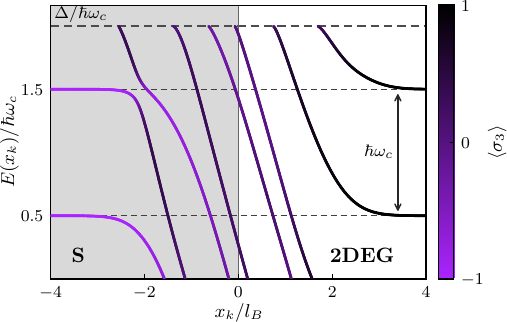}
  \end{center}
  \caption{Spectrum of chiral Andreev edge states. {\color{black}The colour of each branch
encodes the weight of the BdG spinor in the particle and hole sectors, $\langle\sigma_3\rangle=\int dx\,(|f_{n,x_k}|^2
-|g_{n,x_k}|^2)=1-2\langle g_{n,x_k}|g_{n,x_k}\rangle$: purple
($\langle\sigma_3\rangle=-1$) marks purely hole-like and black
($\langle\sigma_3\rangle=+1$) purely electron-like excitations.} The gray dashed line marks the onset of the quasiparticle continuum. {\color{black} The figure corresponds to $\Delta/\omega_c=1, \; \varepsilon_{\rm F}/\omega_c=3 $ (or $2\mathcal{N}=6$)}. The dark white (gray) region corresponds to the 2DEG (S).} 
	\label{Fig_spectrum}
\end{figure}

\noindent\textsf{\color{blue}Clean 2DEG-S.}  
We start with the well-known problem of a clean 2DEG-S interface  
(see Fig.~\ref{Fig_spectrum}) described by  
Bogoliubov-de Gennes (BdG) 
Hamiltonian~\cite{hoppe2000,hoppe2001,giazotto2005}
\begin{gather}
    H_{\rm BdG} =  \frac{\sigma_3}{2m} (-i\hbar \boldsymbol{\nabla}-\sigma_3 e\bm{A}/c)^2+ U_0\delta(x)\sigma_3  - \varepsilon_{\rm F} \sigma_3\notag \\+
    |\Delta| [\cos(2\varphi) \sigma_1+\sin(2\varphi) \sigma_2]\theta(-x) .
     \label{eq:CleanH_BdG}
\end{gather}
Here $\sigma_{1,2,3}$ are standard Pauli matrices acting in the BdG space. The potential
$U(x) = U_0 \delta(x)$ models a non-ideal interface, $|\Delta|$ is the absolute value of the superconducting order parameter, $m$ stands for an effective electron mass, and $\varepsilon_{\rm F}$ is the Fermi energy. Below we will neglect a spatial dependence of $|\Delta|$ due to the proximity effect.  
The vector potential $\bm{A}$ and the superconducting phase $\varphi$ are determined self-consistently from the Maxwell and London equations: $\varphi(y) {=} \varphi_0 {-} y \lambda/l^2_{B}$ and $\bm{A}{=} B [\lambda (e^{x/\lambda}{-}1) \theta({-}x) {+} x \theta(x)] \bm{e}_y$. Here $\lambda$ is the penetration length, $B$ is the magnetic field perpendicular to 2DEG, and $l_{B}= \sqrt{\hbar c / eB}$ is the magnetic length. 

The BdG Hamiltonian \eqref{eq:CleanH_BdG} has no time reversal symmetry but possesses BdG symmetry: 
$H_{\rm BdG}=-\sigma_2 H_{\rm BdG}^T \sigma_2$. In addition, since we do not consider the Zeeman effect, the fermion states corresponding to the Hamiltonian \eqref{eq:CleanH_BdG} are doubly degenerate in spin, i.e. 
$H_{\rm BdG}$ commutes with the spin operator. In disordered systems, a Hamiltonian with such set of symmetries would belong to the class C in Altland-Zirnbauer classification \cite{AltZirnSC}.   

The eigenvalue problem for the BdG Hamiltonian \eqref{eq:CleanH_BdG}, $H_{\rm BdG}\Psi_n = E_n \Psi_n$ can be analysed by standard means under assumption of large enough penetration length (see \cite{SM}). The solution can be written 
as~\cite{giazotto2005} 

\begin{equation}
   \Psi_n(x,y) = \frac{e^{i\sigma_3 \varphi_0}}{\sqrt{L_y}} \begin{pmatrix}
        f_{n, x_k}(x) \\ g_{n, x_k}(x)
    \end{pmatrix} e^{i  y  (\sigma_0 x_k - \sigma_3 \lambda)/l^2_B},
    \label{eq:w.f.}
\end{equation}
where  $x_k = p_y l_B^2$ is the guiding-center coordinate of the skipping orbit in the $x$-direction, $p_y$ is the momentum along the interface, and $L_y$ is the length of the boundary. The eigenfunctions $f_{n, x_k}$, $g_{n, x_k}$ and eigenenergies $E_n>0$ can be found numerically. We focus on the states with energies satisfying $|E_n+\omega_c \lambda x_k/l_B^2|<|\Delta|$ where $\omega_c=eB/(mc)$ is the cyclotron frequency (see Fig.~\ref{Fig_spectrum}). For such states, the eigenfunction $\Psi_n$ describes an evanescent quasiparticle state localized near the interface. Additionally, such sub-gap states contribute only to the supercurrent that flows 
 from 2DEG into the superconductor. 
 
The total quasiparticle current flowing along the boundary 
can be decomposed into three contributions  
(see \cite{SM}): $I_{n,x_k}^{\rm (Q)}=I_{n,x_k}^{\rm(Q,{ N})}-I_{n,x_k}^{\rm (Q,{ A})}+I_{n,x_k}^{\rm(Q,sc)}$, where we distinguish between the ordinary {\iqH} current ($I_{n,x_k}^{\rm(Q,{ N})}$), the Andreev conversion current ($I_{n,x_k}^{\rm(Q,{ A})}$), and the quasiparticle component of the supercurrent ($I_{n,x_k}^{\rm(Q,sc)}$), respectively. The corresponding charge conductance at zero temperature, $T=0$, can be written as follows:
\begin{equation}
G_{\rm Q} = \frac{e^2}{h}\sum_{n=1}^{2\mathcal{N}} \Bigl(1-2\left\langle g_{n,x_{k_0}}|g_{n,x_{k_0}}\right\rangle\Bigr) , 
\label{eq:GQ}
\end{equation}
where $2\mathcal{N} {=} 2\left\lfloor \varepsilon_{\rm F}/\hbar \omega_c \right\rfloor$ is the number of edge modes crossing the Fermi energy and $x_{k_0}$ is the solution of $E_n(x_{k_0}){=}0$. We note that the number of edge modes is always even ($2\mathcal{N}$) due to the BdG symmetry. {\color{black} In addition, the number of topologically protected edge modes can be expressed in terms of the Landau-level filling factor as $\mathcal{N} = \lfloor \nu/2 \rfloor$.} The result~\eqref{eq:GQ} coincides with the results obtained in Refs.~\cite{hoppe2000, hoppe2001, giazotto2005}.

In contrast to the charge current, the total spin current consists of only a single contribution, $I^{\rm(S)}_{n,x_k} \propto  I^{\rm(P)}_{n,x_k}$, where $I^{\rm(P)}$ is the probability current (see \cite{SM}). This proportionality follows from spin conservation for $H_{\rm BdG}$.

 Using a similar formalism, one can obtain the spin conductance as a response of the spin current on the infinitesimal gradient of a Zeeman magnetic field (e.g., created by a Zeeman splitting in a source) or to a spin~bias \footnote{{\color{black} It should not be confused with the spin Hall conductance, i.e., the spin response to an electric field, which characterizes the quantum spin Hall effect in symmetry class AII with preserved time-reversal symmetry and a $\mathbb{Z}_2$, rather than a $2\mathbb{Z}$, topological invariant.}}:
\begin{equation}
G^{\rm (S)} = 2\mathcal{N} \frac{\left(\hbar/2\right)^2}{2\pi \hbar}, \qquad \mathcal{N} \in \mathbb{Z} .
\label{eq:GS}
\end{equation}
  Therefore, the {\color{black} spin } response of the system is quantized in even-integer units of the spin conductance quantum $G^{\rm (S)}_{0} {=} \hbar/8\pi$. This is a hallmark of {\sqH} edge states \cite{Senthil1999}.

\noindent\textsf{\color{blue} Diffusive regime.} In realistic experimental conditions, preparing perfectly clean superconducting samples may be challenging. So, below we show that the integer quantization of the spin conductance, Eq.\eqref{eq:GS}, is independent of impurity scattering, which is the consequence of the topological protection of sqH edge states. We assume that the superconductor is in the dirty limit, $\xi \gg l_{\rm sc}$, where $l_{\rm sc}$ is the mean free path in the superconductor. 
The imaginary-time  
action for the  $\nu=2$ {\iqH} edge modes propagating along the 2DEG-S interface reads:
\begin{equation}
\mathcal{S}_{e} {=} \int\limits_{0}^{\beta} d\tau  \int\limits_{0}^{L} dy_1 dy_2\, 
\hat{\overline{\eta}}(y_1,\tau)
\left[{-}\partial_{\tau}{-}\hat{H}(y_1,y_2)\right]
\hat{\eta}(y_2,\tau) ,
\label{eq:S:e:dis}
\end{equation}
where $\hat{\eta} = \left(\eta_{\uparrow}, -\overline{\eta}_{\downarrow}\right)$ are Grassmann variables corresponding to the chiral edge modes.  
The effective Hamiltonian for these edge modes spreading along the 2DEG-S interface  with a velocity $v$ is non-local in space  
\cite{glazman2023}:
\begin{equation}\label{eq: secdiff: hamdiff}
\hat{H}(y_1,y_2)
=
\delta(y_1 - y_2) \left(-iv \partial_{y_2}\right)
+
\hat{V} (y_1,y_2) .
\end{equation}
The nonlocal matrix random potential $\hat{V}$ can be expressed in terms of the exact $2{\times}2$ superconducting Green’s function $\hat{\mathcal{G}}$ in the presence of impurity scattering.

We approximate $\hat{V}$ as a Gaussian random field with zero mean and a specified two-point correlation function (see Supplementary Material~\cite{SM}):
\begin{gather}
   \left\langle \hat{V}_{\alpha \beta}(y_1,y_2) \hat{V}_{\gamma \delta}(y_3,y_4)\right\rangle= W(|y_{12}|)(2\delta_{\alpha\delta}\delta_{\beta\gamma}{-}\delta_{\alpha\beta}\delta_{\delta\gamma}) \notag\\ \times (\delta(y_{13})\delta(y_{24}){+}\delta(y_{14})\delta(y_{23})) ,
    \label{eq: secdiff: vcorr}
\end{gather}
where $y_{jk}{=}y_j{-}y_k$ and the correlation function 
$W(y){=}t^4 (\partial_x \Phi)^4\pi^2 \nu_F p_F^2 e^{-|y|/\xi}/(2D |y|)$ indicates that spatial correlations extend over distances of the order of the superconducting coherence length $\xi$. Here $\partial_x \Phi(x)$ denotes the derivative of the transverse component of the edge-state wave function evaluated at the interface ($x=0$), and $t$ is the tunneling amplitude. The parameters $\nu_F$, $p_F$, and $D = v_F l_{\rm sc}/3$ correspond to the normal-state density of states, Fermi momentum, and diffusion constant of the superconductor, respectively. Since we are interested in the long-distance (universal) regime, $|y| \gg \xi$, we employ the following approximation:
$W(y)\simeq t^4 (\partial_x \Phi)^4\pi^2(\pi^2 \nu_F p_F^2/D) \delta(y) \ln(\xi/l_{\rm sc})$.

The kinetic part of the Hamiltonian \eqref{eq: secdiff: hamdiff} breaks the time-reversal symmetry, whereas the superconducting correlations associated with the Andreev reflections enforce the BdG symmetry (charge-conjugation) $\hat{H} = -\sigma_2 \hat{H}^T \sigma_2$. Thus a random Hamiltonian 
\eqref{eq: secdiff: hamdiff} belongs to the symmetry class C. Applying standard methods to the action \eqref{eq:S:e:dis}, we derive the {\NLSM} action (see \cite{SM}): 
\begin{equation}
\mathcal{S}[Q] {=} \frac{\mathcal{N}}{2}\int\limits_0^L dy  \operatorname{Tr} \Lambda T\partial_y T^{-1} {-}\frac{\mathcal{N} \ell_A}{8} \int\limits_0^L dy \operatorname{Tr} (\partial_y Q)^2 .
\label{eq:NLSM:1D}
\end{equation}
Here a Hermitian matrix $Q{=}T^{{-}1}\Lambda T$ acts in the $N_r{\times}N_r$ replica space, 
$2N_m{\times}2N_m$ Matsubara space with fermionic frequencies $\varepsilon_n{=}\pi T(2n{+}1)$, and 
$2{\times}2$ Nambu space spanned by the Pauli matrices $s_{1,2,3}$ and the unit matrix $s_0$.
The matrix $\Lambda_{\varepsilon_n,\varepsilon_m} {=} \operatorname{sgn}({\varepsilon_n})\hat{1}_r s_0$
describes the standard metallic saddle-point.

The matrix $Q$ satisfies the charge-conjugation constraint: $Q {=} {-}L_0 s_2 Q^T s_2 L_0$, where $\left(L_0\right)_{\varepsilon_n,\varepsilon_m}{=} \delta_{\varepsilon_n, -\varepsilon_m}\hat{1}_r s_0$, that restricts the {\NLSM} manifold to $Q\in \mathrm{Sp}(4N_m N_r)/\mathrm{U}(2N_m N_r)$.  In derivation of the {\NLSM} action we take into account the existence of $2\mathcal{N}$ edge channels near the boundary as considerations in the clean case demonstrates~\footnote{{ Technically, derivation of the {\NLSM} action \eqref{eq:NLSM:1D} requires the condition $\mathcal{N}\gg 1$. However, since the structure of the {\NLSM} action is dictated by the symmetries it is believed that the action \eqref{eq:NLSM:1D} holds for all $\mathcal{N}$.}}. The quantity $1/\ell_A$ plays a role of 
the inverse effective mean free path for the (Andreev) edge states. For $\mathcal{N}{=}1$ we find
that $1/\ell_A = 2 \pi^2 t^4 \left(\partial_x \Phi\right)^4 (\nu_F p_F^2 /v^2 D) \ln(\xi/l_{\rm sc})$ 
in agreement with the expression derived in Ref. \cite{glazman2023}. \color{black}
Therefore, we conclude that the edge states propagating along the interface between 2DEG and a dirty superconductor are described by the {\NLSM} for the class C, Eq. \eqref{eq:NLSM:1D}, i.e., they are nothing but the {\sqH} edge states. 

For a long boundary, $L \gg \xi$, the spin conductance remains  quantized with magnitude $ G^{\rm (S)} = 2\mathcal{N} G^{\rm (S)}_{0}$ \cite{Read2000, Senthil1999}, in agreement with the results of the previous section, cf. Eq. \eqref{eq:GS}. Therefore, we conclude that {\sqH} edge modes are topologically protected against disorder scattering.  

Observation of the even-integer quantized spin Hall conductance is experimentally challenging. Measuring a spin current response to a nonuniform Zeeman splitting  requires a control over the edge states spectrum via a spatially nonuniform $g$-factor in the 2DEG~\cite{Salis2001} or via a ferromagnetic proximity effect~\cite{Zhao1995,fominov2017}, accompanied by a measurement of magnetic moment of the edge currents. In the following, we propose a feasible experimental scheme capable of measuring the quantized conductance of {\sqH} edge states in a hybrid 2DEG-S system in the quantum Hall regime at a filling factor $\nu=2$.

{\noindent\textsf{\color{blue} Possible experimental setup.} 
Our proposal exploits that a weak Zeeman field leaves the edge spin conductance quantized and topologically robust \cite{Parfenov2025}.
 
In the presence of Zeeman splitting, the  2DEG in the  $\nu=2$ quantum Hall regime 
supports two spin-split chiral edge channels ({\ECs}) and enables individual control over their chemical potentials and currents by all-electrical means~\cite{vanWees1991,Beenakker1991,Carrega2021,Zimmermann2017}. The sketch of the proposed device is shown in Fig.~\ref{Fig_experimental}a. The thick black lines mark the boundary of the 2DEG, along which the {\ECs} propagate in a clockwise direction (see arrows). Two gate-voltage-defined constrictions are used to manipulate the {\ECs}~\footnote{{\color{black} Alternatively, one may use a single narrow top gate tuned to a local filling factor $\nu=1$ in spirit of Ref.~\cite{matsuo2018equal}, which is the easiest way to reach a 100\% spin selectivity.}}. Each of them is tuned to fully transmit the inner (spin-down) {\ECs} and fully reflect the outer (spin-up) one~\footnote{{\color{black}Such sequence of inner and outer {\ECs} assumes negative $g$-factor of 2DEG.}}.
 
Downstream of the right constriction, the {\ECs} acquire chemical potentials $\mu_\downarrow$ and $\mu_\uparrow$, respectively, which can be tuned {\color{black} independently} by the bias voltages applied to the normal terminals upstream of the right constriction{\color{black}, mimicking the spin bias}. Consider the case of the spin-up chemical potential slightly higher than that of the spin-down one, and both are above the chemical potential of the grounded superconductor $\mu_\uparrow>\mu_\downarrow>\mu_S=0$. The corresponding equilibrium electronic energy distributions $f(E)$ before the {\ECs} reach the 2DEG-S boundary are illustrated in the inset i) of Fig.~\ref{Fig_experimental}a (right). As the {\ECs} further propagate along the proximitized 2DEG-S boundary the quasiparticles experience Andreev reflection with a probability $\textsf{T}_{\rm A}$. The resulting non-equilibrium energy distributions downstream of the superconducting terminal are determined by $\textsf{T}_{\rm A}$. Note that the Andreev reflection process simultaneously removes a spin-up electron with energy $E$ and a spin-down electron with energy $-E$ (energy is counted from $\mu_S$). Hence, the energy distributions for the spin-up and spin-down {\ECs} downstream of the superconductor are mutually correlated, as shown in the left inset i) of Fig.~\ref{Fig_experimental}a. {\color{black} We note that, unlike the charge equilibration discussed in Ref.~\cite{Wen2018}, the spin {\color{black} bias} cannot relax into a singlet superconducting condensate.} The left constriction splits the {\ECs} and enables a separate measurement of the spin-up, $I_\uparrow$, and spin-down, $I_\downarrow$, currents at the two other normal terminals downstream. 

\begin{figure}[t!]
\begin{center}
\vspace{0mm}
 \includegraphics[width=1\linewidth]{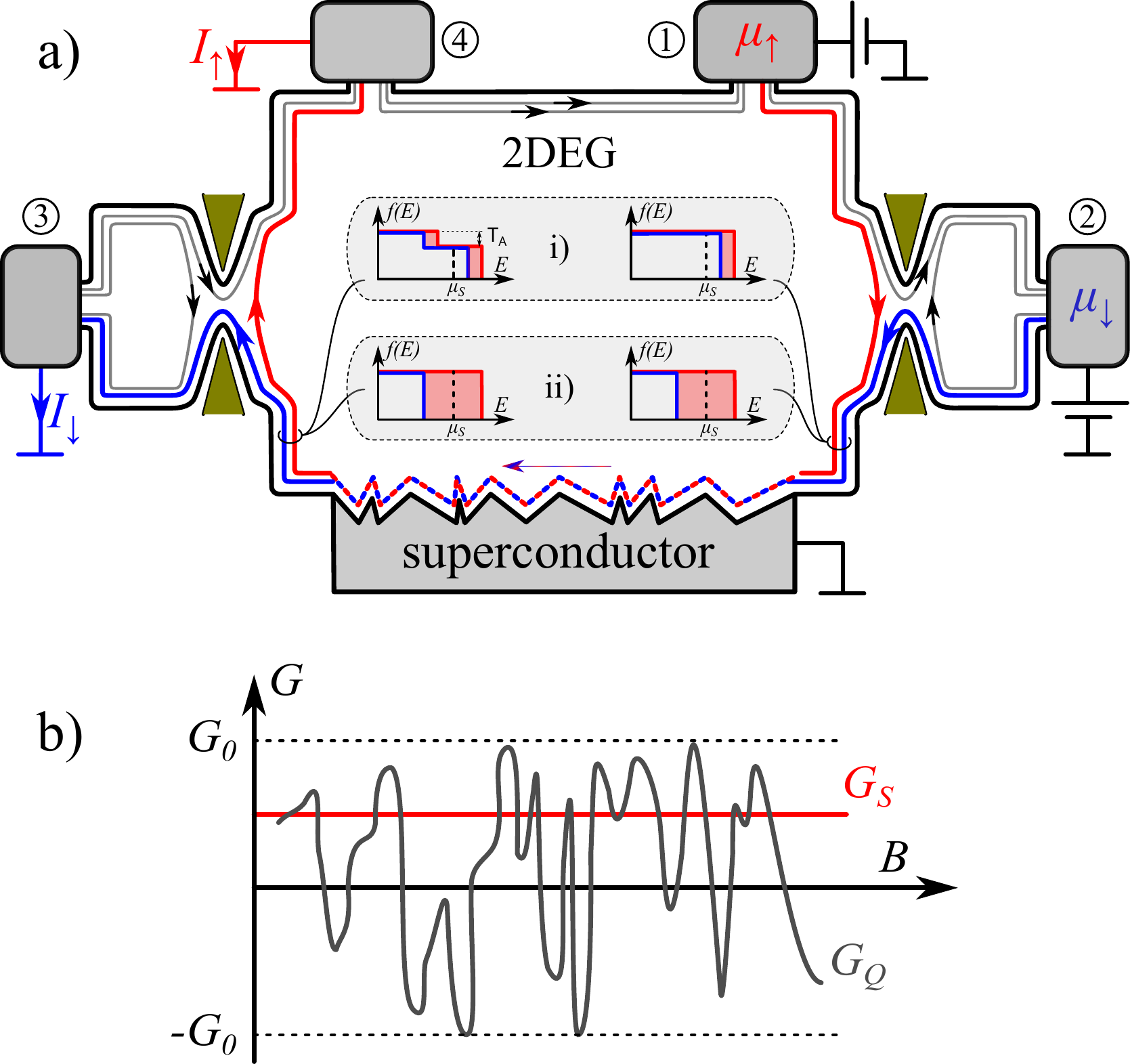}
  \end{center}
  \caption{Possible experimental realization. (a): Schematic sketch of the proposed device, based on the 2DEG-S hybrid system in the 
  {\iqH} regime at a factor $\nu=2$. Two gate-voltage-defined constrictions transmit the inner EC and reflect the outer EC, thereby enabling separate biasing and current measurement of the spin species. Insets: electronic energy distributions before the EC enter the superconducting proximity region and after they leave it. Inset i) corresponds to the case of $\mu_\uparrow>\mu_\downarrow>\mu_S$. Inset ii) is for $(\mu_\uparrow+\mu_\downarrow)/2=\mu_S$. {\color{black} 
  Red (blue) color marks the spin-up/outer (spin-down/inner) EC with chemical potential $\mu_\uparrow$ ($\mu_\downarrow$) and current $I_\uparrow$ ($I_\downarrow$); the same colour code is used in the insets.
  } 
  (b): Schematic dependence of the charge and spin conductances on the $B$-field. We predict that, unlike the strongly fluctuating charge conductance $G_{\rm Q}$, the spin conductance $G_{\rm S}$ remains integer quantized due to topological protection (see text). } 
	\label{Fig_experimental}
\end{figure}

We analyze charge and spin transport in this system using the Landauer--B\"uttiker formalism generalized to superconducting systems \cite{Anantram1996,Michelsen2023}. An electron/hole contribution to the current in the $i$-th terminal is given by 
\begin{equation}
    I^{\alpha}_{i} = \frac{\sigma_{\alpha} e}{2h} \sum_{\beta=e,h}\sum_{j}\int d\varepsilon \left[ \textsf{T}^{\alpha\beta}_{ji}f^{\beta,i}_{\rm F}(\varepsilon)  {-} \textsf{T}^{\alpha\beta}_{ij}f^{\beta,j}_{\rm F}(\varepsilon)\right] .
    \label{eq:qcurrent}
\end{equation}
Here, $f^{\beta,j}_{\rm F}(\varepsilon)=f_{\rm F}(\varepsilon-\sigma_{\beta}\mu_j)$, where $\sigma_{e/h} = \pm 1$, is the Fermi-Dirac distribution function in the $j$-th terminal.  Also, we define $\textsf{T}^{\alpha\beta}_{ij}$ as the transmission probability for a quasiparticle of type $\beta$ incident from  
the $j$-th lead to be transmitted to  
the $i$-th lead as a quasiparticle of type $\alpha$.
 They can be calculated using the transfer-matrix formalism as $\textsf{T}^{ee/hh}_{ij}  = \textsf{T}_{\rm N}$ and  $\textsf{T}^{eh/he}_{ij} =\textsf{T}_{\rm A}$ where $\textsf{T}_{\rm N}=1-\textsf{T}_{\rm A}$ is the probability of the normal transmission (see \cite{SM}). 
For the setup shown in Fig.~\ref{Fig_experimental}a, {\color{black} and assuming perfect spin filtering at the constrictions}, the currents $I_{\downarrow,\uparrow}=I^{e}_{3,4}+I^{h}_{3,4}$ are obtained from Eq.~\eqref{eq:qcurrent} (see the \cite{SM} for details). {\color{black}Using this result, we obtain the spin current $I_{\rm S}=I_\uparrow-I_\downarrow$:
\begin{equation}\label{eq:IQS:cur}
    I_{\rm S} {=} \frac{e}{2h} \int d{\varepsilon} \left [\mathcal{F}(\varepsilon,\mu_{\uparrow}) {-}\mathcal{F}(\varepsilon,\mu_{\downarrow})\right]{=} G_{\rm S}\frac{\left(\mu_\uparrow{-}\mu_\downarrow\right)}{e} ,
\end{equation}
where $\mathcal{F}(\varepsilon,\mu_i)=f_{\rm F}(\varepsilon-\mu_i)-f_{\rm F}(\varepsilon+\mu_i)$. Here $G_{\rm S}$ denotes the spin conductance (here we use a different notation for the spin conductance than in Eq.~\eqref{eq:GS} because of its different dimension). It is independent of $\textsf{T}_{\rm A}$ and topologically protected against mesoscopic fluctuations, taking the quantized value (Fig.~\ref{Fig_experimental}b)
\begin{equation}
G_{\rm S}=G_0/2=e^2/h .
\label{eq:GS:e}
\end{equation}
Notably, Eq.~\eqref{eq:IQS:cur} and the corresponding quantization of $G_{\rm S}$ are universal, independent of the microscopic model of the interface, and remain valid beyond linear response, provided transport stays in the subgap regime.

By contrast, for the charge current $I_{\rm Q}=I_\uparrow+I_\downarrow$ we find $I_{\rm Q}=G_{\rm Q}(\mu_\uparrow+\mu_\downarrow)/(2e)$, where $G_{\rm Q}=(1-2\textsf{T}_{\rm A})G_0$  {\color{black} in the linear response regime}. {\color{black} As shown in Ref.~\cite{glazman2023}, in the presence of disorder the $\textsf{T}_{\rm A}$ fluctuates and the $G_{\rm Q}$ exhibits mesoscopic fluctuations as a function of the interface length, carrier density, or magnetic field, as illustrated in Fig.~\ref{Fig_experimental}b.  Unlike the spin conductance, the $G_{\rm Q}$ averages out  beyond the linear response regime.}  

} 

We emphasize that the result in Eq.~\eqref{eq:GS:e} is directly related to the even-integer quantized spin conductance in Eq.~\eqref{eq:GS}. Indeed, consider   
a fine-tuned situation of $\mu_{\uparrow} = -\mu_{\downarrow}$ (mimicking Zeeman splitting), for which the phase space for Andreev reflection vanishes since a spin-up electron has no corresponding spin-down partner~\footnote{Although in describing the experimental setup we assumed $\mu_{\uparrow} > \mu_{\downarrow} > 0$, the expressions in Eq.~\eqref{eq:IQS:cur} remain valid for arbitrary relations between $\mu_{\uparrow}$ and $\mu_{\downarrow}$.}. In this case, the energy distributions upstream and downstream of the superconductor coincide, see the inset ii) of Fig.~\ref{Fig_experimental}a,
and  the charge current vanishes, $I_{\rm Q} \equiv 0$. At the same time, the spin current reduces to $I_{\rm S} = 2G_{\rm S} \mu_{\uparrow}$, corresponding to the even-integer quantized spin conductance given in Eq.~\eqref{eq:GS}.

Randomness of Andreev reflections gives rise to spontaneous fluctuations in both spin-up and spin-down currents {\color{black} simultaneously}. As a consequence, the momentary fluctuations in spin-split edge channels are identical, and the spin currents $I_{\uparrow}$ and $I_{\downarrow}$ are fully cross-correlated: $\langle\delta I_\downarrow^2\rangle = \langle\delta I_\downarrow \delta I_\uparrow\rangle = \langle\delta I_\uparrow^2\rangle$, reflecting the topological protection of the ECs and the spin current $I_S$. These correlations can be observed even {\color{black} without the spin bias for} $\mu_\downarrow = \mu_\uparrow$, i.e., when our setup resembles a Cooper pair splitter. Note that the use of a spin-selective beam splitter results in unit efficiency of Cooper pair splitting, unlike setups with spin-nonselective beam splitters~\cite{Torres1999,Lesovik2001,Khrapai2023}, which are limited to 50\% efficiency~\cite{Tikhonov2024}.
\color{black}

\color{black}

\noindent\textsf{\color{blue}Discussions and conclusions.} The {\sqH} edge states are known to exhibit topologically protected thermal transport with quantized thermal conductance \cite{Senthil1999}. We therefore suggest that the topological protection of {\sqH} edge states in 2DEG-S hybrid structures can be also probed via thermal transport measurements \cite{Zhao2025}. 

{\color{black} In sqH effect, the spin conservation plays the same role as the absence of backscattering in the iqH effect~\cite{Klitzing1980,Tsui1981,Roulleau2008}, therefore \color{black} quantization of the spin conductance does not rely on phase coherence along the 2DEG--S interface. {\color{black} Inelastic scattering brings the electronic energy distributions of the spin resolved ECs towards local equilibrium~\cite{Wen2018}, but leaves the  spin current derived in Eq.~(\ref{eq:IQS:cur}) unchanged. }



In contrast, spin-relaxation mechanisms may affect the quantization by breaking spin-rotation symmetry. Still the integer quantization of the spin conductance in a 2DEG--S hybrid structure survives provided $L\ll\ell_{\rm s}$, where $\ell_{\rm s}$ is the spin-relaxation length. 
This regime is achievable in typical semiconductor- and graphene-based 2DEGs.
For $L\gtrsim\ell_{\rm s}$, the quantization is destroyed 
and the system crosses over from class C to class D.} In two dimensions, class D possesses an integer-valued topological invariant \cite{Kitaev2009,Schnyder2009} responsible for the thermal quantum Hall effect. Hence, the edge states remain topologically protected. This protection can likewise be probed via thermal transport. Other signatures of topological protection of the class D edge states in 2DEG-S hybrids with spin-orbit coupling have recently been investigated in Ref.~\cite{Baba2026}. 

 Another mechanism that can destroy the quantization of spin conductance is the presence of Abrikosov vortices near the 2DEG-S interface. In this case, edge quasiparticles can tunnel into vortex-core states, thereby being removed from edge transport and suppressing both charge and spin conductances. We note that the effect of vortices can be studied similarly to Refs.~\cite{antonenko2024, antonenko2025}, where a 2DEG in a perpendicular magnetic field was considered, with a superconducting order parameter corresponding to a vortex lattice; an even-integer bulk topological invariant, as well as chiral edge modes, were identified in those works. A detailed analysis of the effect of vortices is beyond the scope of the present work.

In experimental setups where a narrow superconducting bridge is embedded in a 2DEG, crossed Andreev reflection processes can occur \cite{Lee2017,glazman2023ii,lee2022}. Such processes may suppress the topological protection of the {\sqH} edge modes by enabling backscattering mediated by crossed Andreev reflection. We note that the universal localization theory for such counterpropagating modes was developed in Ref.~\cite{khalafthes,khalaf2016}.

To summarize, we argue that chiral Andreev edge states propagating along the interface between a superconductor proximity-coupled to a 2DEG in a magnetic field are nothing but the edge states of the spin quantum Hall effect. Remarkably, this observation explains why the charge transport is strongly affected by disorder. In contrast, we predict that the spin transport remains robust and exhibits even-integer quantization. Our results thus establish a concrete route to realizing and detecting spin quantum Hall edge states in time-reversal-symmetry-breaking superconducting hybrids. We suggest how the topological protection of these edge states can be probed via standard electrical measurements.

\noindent\textsf{\color{blue}Acknowledgements} We thank 
Y. Fominov, P. Ostrovsky, and  A. Mel'nikov for enlightening discussions. One of us (I.S.B.) is grateful to I. Gornyi, I. Gruzberg, and H. Obuse for past collaboration on the related project that partially motivated this work. This work was partially supported by the Russian Science Foundation under Grant No. 25-42-01036 (research on the diffusive regime) and by the
Ministry of Science and Higher Education of the Russian
Federation under Projects No. FFWR-2024-0017 (research on the ballistic regime). The authors acknowledge personal support from the Foundation for the Advancement of Theoretical Physics and Mathematics ``BASIS''. M.V.P. is grateful to ICQM, Peking University for hospitality.

\noindent\textsf{\color{blue}Data Availability} The data that support the findings of this article are not publicly available. The data are available from the authors upon reasonable request.

\bibliography{literature_classC_top}



\section*{Supplementary material}

# ONLINE SUPPORTING INFORMATION
# Emergence of spin quantum Hall edge states at the boundary of two-dimensional electron gas proximized with an $s$-wave superconductor

M. V. Parfenov,[1,2,3] V. S. Khrapai,[4,5] and I. S. Burmistrov[1,2,3]

[1] *L. D. Landau Institute for Theoretical Physics, Semenova 1-a, 142432, Chernogolovka, Russia*
[2] *Laboratory for Condensed Matter Physics, HSE University, 101000 Moscow, Russia*
[3] *P. N. Lebedev Physical Institute, Russian Academy of Sciences, 119991 Moscow, Russia*
[4] *Osipyan Institute of Solid State Physics, Russian Academy of Sciences, 142432 Chernogolovka, Russian Federation*
[5] *HSE University, 101000 Moscow, Russia*

In this notes we present details of (i) solution of the Bogoliubov–de Gennes equation at the S-2DEG interface, (ii) edge current derivation, (iii) edge NL$\sigma$M derivation and (iv) transfer-matrix approach for transmission amplitudes.

## S.I. SOLUTION OF THE BOGOLIUBOV–DE GENNES EQUATION AT THE S-2DEG INTERFACE

In this section we show the detailed solution for the BdG Hamiltonian for the 2DEG-S inteface. This problem was previously studied in Refs. [1–3]. The Hamiltonian of this system takes the form

$$\begin{pmatrix} H_0 + U - \varepsilon_{\rm F} & |\Delta| e^{2i\varphi} \\ |\Delta| e^{-2i\varphi} & -H_0^T - U + \varepsilon_{\rm F} \end{pmatrix} \begin{pmatrix} u \\ v \end{pmatrix} = E \begin{pmatrix} u \\ v \end{pmatrix}. \tag{S.1}$$

Here $H_0$ denotes the quasiparticle Hamiltonian (different in the S and N regions), $U(x) = U_0\delta(x)$ is a potential that models a non-ideal interface, $\Delta = |\Delta| e^{2i\varphi}\theta(-x)$ is the superconducting pairing potential and $\varepsilon_{\rm F}$ is the Fermi energy. Here we consider a superconducting layer with a finite penetration depth $\lambda$. In this case, the magnetic field $\boldsymbol{B} = B\boldsymbol{e}_z$ does not vanish inside the superconductor but decays exponentially (Meissner effect):

$$B(x) = \Big( B e^{x/\lambda}\theta(-x) + B\theta(x) \Big), \quad A(x) = \lambda B \left( e^{x\lambda} - 1 \right)\theta(-x) + xB\theta(x), \tag{S.2}$$

where $\theta(x)$ is Heaviside step function. Due to the presence of non-zero vector potential in superconducting region, the pairing potential phase $\varphi$ is non-uniform and must therefore be determined self-consistently from the Maxwell and London equations (see Refs. [1, 3] for details). From the second London equation supercurrent can be written in form [4]:

$$j_s^{(y)} = \frac{c}{4\pi\lambda^2} \left( \frac{\hbar c}{e}\partial_y \varphi - A(x) \right). \tag{S.3}$$

Meanwhile supercurrent can be found from Maxwell equation:

$$\operatorname{rot} \boldsymbol{B} = \frac{4\pi}{c}\boldsymbol{j}_s \quad \Rightarrow \quad \boldsymbol{j}_s = -\frac{cB}{4\pi\lambda} e^{x/\lambda}. \tag{S.4}$$

After substituting this result to the London equation, one finds

$$\varphi(y) = \varphi_0 - \frac{y\lambda}{l_B^2}, \tag{S.5}$$

where $l_B^2 = \hbar c/eB$ is the magnetic length. It is convenient to gauge out the phase of the superconducting order parameter and transfer it into the phase of the BdG spinors. In the normal region, the quasiparticle Hamiltonian takes the form of the standard Pauli Hamiltonian:

$$H_0^{(N)} = \frac{p_x^2}{2m} + \frac{m\omega_c^2}{2} \left( x - (x_k - \lambda) \right)^2, \tag{S.6}$$

where $x_k = p_y l_B^2$ is the guiding-center coordinate of the skipping orbit in the $x$-direction. For holes, one should replace $x_k \to -x_k$. In the superconducting region the Hamiltonian becomes

$$H_0^{(S)} = \frac{p_x^2}{2m} + \frac{m\omega_c^2}{2} \left( x_k - \lambda \right)^2. \tag{S.7}$$

We note that a finite penetration depth $\lambda$ shifts the guiding-center coordinate.

The solution of the Bogoliubov–de Gennes equation (S.1) can be written in the form (see Ref. [1])

$$\begin{pmatrix} u \\ v \end{pmatrix} = \frac{e^{i\sigma_3\varphi_0}}{\sqrt{L_y}} \begin{pmatrix} f_{n,x_k}(x) \\ g_{n,x_k}(x) \end{pmatrix} e^{iy(\sigma_0 x_k - \sigma_3\lambda)/l_B^2}, \tag{S.8}$$

where $s_i$ are the Pauli matrices in the BdG space. In our analysis we neglect particle motion along the $z$-direction, since it leads only to a trivial renormalization of the Fermi energy $\varepsilon_F$. For $x < 0$ (the superconducting region), the functions $f, g$ can be expressed as plane waves:

$$\begin{pmatrix} f_{n,x_k}(x) \\ g_{n,x_k}(x) \end{pmatrix}_{x<0} = C_1 \begin{pmatrix} \gamma_- \\ 1 \end{pmatrix} e^{-ixq_-} + C_2 \begin{pmatrix} \gamma_+ \\ 1 \end{pmatrix} e^{ixq_+} \tag{S.9}$$

with non-trivial parameters determined by solution of the corresponding BdG equation:

$$q_\pm = \left( 2m \left( \varepsilon_{\rm F} \pm \sqrt{E_\lambda^2 - |\Delta|^2} \right) - \frac{x_k^2 + \lambda^2}{l_B^4} \right)^{1/2}, \quad \gamma_\pm = \frac{|\Delta|}{E_\lambda \mp \sqrt{E_\lambda^2 - |\Delta|^2}}, \quad E_\lambda = E + \hbar\omega_c \frac{\lambda x_k}{l_B^2}. \tag{S.10}$$

Here $E_\lambda$ denotes the Doppler-shifted quasiparticle energy. We restrict our consideration to sub-gap excitations satisfying $|E_\lambda| < |\Delta|$. In this case, the signs of the plane-wave components determine the evanescent quasiparticle states localized near the interface. Within this approximation, no quasiparticle current flows in the superconductor perpendicular to the interface.

For the normal region $x > 0$ solution can be expressed in terms of parabolic cylinder functions $\mathcal{D}_\nu(z)$:

$$\begin{pmatrix} f_{n,x_k} \\ g_{n,x_k} \end{pmatrix}_{x>0} = \sum_{p=\pm} A_p \begin{pmatrix} 1+p \\ 1-p \end{pmatrix} \mathcal{D}_{\frac{pE-\varepsilon_{\rm F}}{\hbar\omega_c} - \frac{1}{2}}(\sqrt{2}\xi_p) \tag{S.11}$$

where $\xi_p = (x + \lambda \mp x_k)/l_B$ and $\mathcal{D}_\nu(z)$ is the parabolic cylinder (Weber-Hermite) function [5]

The corresponding BdG wave functions must be continuous at $x = 0$ and exhibit a discontinuity in their derivatives due to the presence of the interfacial potential barrier $U(x)$. As usual, the energy spectrum $E(x_k)$ of the system can be obtained from the condition that the determinant of the homogeneous linear system for the coefficients $C_1$, $C_2$, $A$ and $B$ vanishes. The numerical solution of the secular equation for the energy spectrum is shown in Fig.(1) in the main text.

## S.II. DERIVATION OF SPIN AND CHARGE CURRENTS

To compute spin and charge currents, we employ microscopic operators in the second-quantized form:

$$\hat{j}_y^{\rm Q} = \frac{e}{2} \sum_{\sigma=\uparrow,\downarrow} \left( \hat{\psi}_\sigma^\dagger \hat{v}_y \hat{\psi}_\sigma + \hat{v}_y^* \hat{\psi}_\sigma^\dagger \hat{\psi}_\sigma \right), \quad \hat{j}_y^{\rm S_z} = \frac{\hbar/2}{2} {\rm tr}_\sigma \left( \hat{\psi}^\dagger \sigma_z \hat{v}_y \hat{\psi} + \hat{v}_y^* \sigma_z \hat{\psi}^\dagger \hat{\psi} \right) \tag{S.12}$$

with the velocity operator defined as $m\hat{v}_y = \hat{p}_y - \frac{e}{c}\hat{A}_y$. We express the fermionic field operators in the Bogoliubov basis:

$$\hat{\psi}_\sigma = \sum_{n,k} \left( u_{n,k,\sigma} \hat{\gamma}_{n,k,\sigma} - \sigma v_{n,k,\sigma}^* \hat{\gamma}_{n,k,-\sigma}^\dagger \right), \quad \hat{\psi}_\sigma^\dagger = \sum_{n,k} \left( u_{n,k,\sigma}^* \hat{\gamma}_{n,k,\sigma}^\dagger - \sigma v_{n,k,\sigma} \hat{\gamma}_{n,k,-\sigma} \right) \tag{S.13}$$

where $u$ and $v$ are solutions of the Bogoliubov–de Gennes equations, and $\hat{\gamma}$ ($\hat{\gamma}^\dagger$) are quasiparticle annihilation (creation) operators. Their expectation values are given by $\left\langle \hat{\gamma}_{n_1,k_1,\sigma_1}^\dagger \hat{\gamma}_{n,k,\sigma} \right\rangle = \delta_{n,n_1}\delta_{k,k_1}\delta_{\sigma,\sigma_1} f_{\rm F}(E_{n,k,\sigma})$, where $f_{\rm F}$ is the Fermi–Dirac distribution. With the help of this we can compute the average over Fermi statistics. Evaluating the averages, we obtain the current densities:

$$\left\langle \hat{j}_y^{\rm Q} \right\rangle = e \sum_{n,k,\sigma} {\rm Re} \left[ u_{n,k,\sigma}^* \hat{v}_y u_{n,k,\sigma} f_{\rm F}(E_{n,k,\sigma}) + v_{n,k,\sigma} \hat{v}_y v_{n,k,\sigma}^* f_{\rm F}(-E_{n,k,-\sigma}) \right] \tag{S.14}$$

$$\left\langle \hat{j}_y^{\rm S_z} \right\rangle = \frac{\hbar}{2} \sum_{n,k,\sigma} {\rm Re} \left[ u_{n,k,\sigma}^* \sigma \hat{v}_y u_{n,k,\sigma} f_{\rm F}(E_{n,k,\sigma}) + v_{n,k,\sigma} \sigma \hat{v}_y v_{n,k,\sigma}^* f_{\rm F}(-E_{n,k,-\sigma}) \right] \tag{S.15}$$

The total current is obtained by integrating over the transverse coordinate $x$. Using the explicit form of the BdG Hamiltonian, we find

$$\int dx\,\mathrm{Re}\left(u^*_{n,k,\sigma}\sigma\hat{v}_y u_{n,k,\sigma}\right) = \frac{\sigma}{\hbar}\frac{l_B^2}{L_y}\left\langle f_{n,\sigma,x_k}\left|\frac{\partial H^{(e)}_{\mathrm{BdG}}}{\partial x_k}\right|f_{n,\sigma,x_k}\right\rangle,$$
$$\int dx\,\mathrm{Re}\left(v_{n,k,\sigma}\sigma\hat{v}_y v^*_{n,k,\sigma}\right) = \frac{\sigma}{\hbar}\frac{l_B^2}{L_y}\left\langle g_{n,\sigma,x_k}\left|-\frac{\partial H^{(h)}_{\mathrm{BdG}}}{\partial x_k}\right|g_{n,\sigma,x_k}\right\rangle \tag{S.16}$$

where $H^{(e/h)}_{\mathrm{BdG}} = \mathrm{tr}[(1\pm\sigma_3)H_{\mathrm{BdG}}/2]$ is the electron/hole sector of the BdG Hamiltonian [Eq. (S.1)]. The resulting charge and spin currents read

$$I^{(\mathrm{Q})} = \frac{e}{\hbar}\frac{l_B^2}{L_y}\sum_{n,k,\sigma}\left(\left\langle f_{n,\sigma,x_k}\left|\frac{\partial H^{(e)}_{\mathrm{BdG}}}{\partial x_k}\right|f_{n,\sigma,x_k}\right\rangle f_{\mathrm{F}}\left(E_{n,k,\sigma}\right) + \left\langle g_{n,\sigma,x_k}\left|-\frac{\partial H^{(h)}_{\mathrm{BdG}}}{\partial x_k}\right|g_{n,\sigma,x_k}\right\rangle f_{\mathrm{F}}\left(-E_{n,k,-\sigma}\right)\right) \tag{S.17}$$

$$I^{(\mathrm{S})} = \frac{\hbar/2}{\hbar}\frac{l_B^2}{L_y}\sum_{n,k,\sigma}\sigma\left(\left\langle f_{n,\sigma,x_k}\left|\frac{\partial H^{(e)}_{\mathrm{BdG}}}{\partial x_k}\right|f_{n,\sigma,x_k}\right\rangle f_{\mathrm{F}}\left(E_{n,k,\sigma}\right) + \left\langle g_{n,\sigma,x_k}\left|-\frac{\partial H^{(h)}_{\mathrm{BdG}}}{\partial x_k}\right|g_{n,\sigma,x_k}\right\rangle f_{\mathrm{F}}\left(-E_{n,k,-\sigma}\right)\right) \tag{S.18}$$

In the low-bias regime, the conductances are obtained as

$$G_{\mathrm{Q}} = -\left.\frac{\partial I^{(\mathrm{Q})}}{\partial V}\right|_{V=0},\quad G_{\mathrm{S}} = -\left.\frac{\partial I^{(\mathrm{S})}}{\partial V_{\mathrm{S}}}\right|_{V_S=0},\quad f_{\mathrm{F}}\left(E_{n,k,\sigma}\right)\to f_{\mathrm{F}}\left(E_{n,k,\sigma}-eV-\sigma g\mu_B B/2\right) \tag{S.19}$$

The spin bias is defined via $\hbar V_{\mathrm{S}}/2 = g\mu_B B/2$. The sign convention corresponds to the direction of the current. In the linear-response regime, we obtain

$$I^{(\mathrm{Q})} = V\frac{2e^2}{\hbar}\frac{l_B^2}{L_y}\sum_{n,k}\left(-\frac{\partial f_{\mathrm{F}}\left(E_{n,k}\right)}{\partial E_{n,k}}\right)\left(\left\langle f_{n,x_k}\left|\frac{\partial H^{(e)}_{\mathrm{BdG}}}{\partial x_k}\right|f_{n,x_k}\right\rangle - \left\langle g_{n,x_k}\left|-\frac{\partial H^{(h)}_{\mathrm{BdG}}}{\partial x_k}\right|g_{n,x_k}\right\rangle\right) \tag{S.20}$$

$$I^{(\mathrm{S})} = V_{\mathrm{S}}\frac{2(\hbar/2)^2}{\hbar}\frac{l_B^2}{L_y}\sum_{n,k}\left(-\frac{\partial f_{\mathrm{F}}\left(E_{n,k}\right)}{\partial E_{n,k}}\right)\left(\left\langle f_{n,x_k}\left|\frac{\partial H^{(e)}_{\mathrm{BdG}}}{\partial x_k}\right|f_{n,x_k}\right\rangle + \left\langle g_{n,x_k}\left|-\frac{\partial H^{(h)}_{\mathrm{BdG}}}{\partial x_k}\right|g_{n,x_k}\right\rangle\right) \tag{S.21}$$

Here we used that, in the absence of spin-dependent terms in the BdG Hamiltonian, the spectrum is doubly degenerate. The corresponding matrix elements can be evaluated using a generalized Hellmann–Feynman theorem:

$$\left\langle f_{n,x_k}\left|\frac{\partial H^{(e)}_{\mathrm{BdG}}}{\partial x_k}\right|f_{n,x_k}\right\rangle + \left\langle g_{n,x_k}\left|-\frac{\partial H^{(h)}_{\mathrm{BdG}}}{\partial x_k}\right|g_{n,x_k}\right\rangle = \frac{\partial E_{n,k}}{\partial x_k} \tag{S.22}$$

$$\left\langle f_{n,x_k}\left|\frac{\partial H^{(e)}_{\mathrm{BdG}}}{\partial x_k}\right|f_{n,x_k}\right\rangle - \left\langle g_{n,x_k}\left|-\frac{\partial H^{(h)}_{\mathrm{BdG}}}{\partial x_k}\right|g_{n,x_k}\right\rangle = \frac{\partial E_{n,k}}{\partial x_k}\left(1-2\left\langle g_{n,x_k}|g_{n,x_k}\right\rangle\right)$$
$$+\,2|\Delta|\left(\langle g_{n,x_k}|\theta(-x)|\partial_x f_{n,x_k}\rangle - \langle f_{n,x_k}|\theta(-x)|\partial_x g_{n,x_k}\rangle\right) \tag{S.23}$$

where we used the normalization of the BdG spinor. The expression for the charge current was previously derived in Ref. [2]. Here, we explicitly compute the spin conductance. Using Eq. (S.22), we find

$$I^{(\mathrm{S})} = V_{\mathrm{S}}\frac{2(\hbar/2)^2}{\hbar}\frac{l_B^2}{L_y}\sum_n\int\frac{L_y dk}{2\pi}\left(-\frac{\partial f_{\mathrm{F}}\left(E_{n,k}\right)}{\partial E_{n,k}}\right)\frac{\partial E_{n,k}}{\partial x_k} = -V_{\mathrm{S}}\frac{(\hbar/2)^2}{2\pi\hbar}2\mathcal{N} \tag{S.24}$$

where $2\mathcal{N}$ is the number of energy levels crossing zero energy, in agreement with Eq. (4) of the main text.

## S.III. EDGE NLσM DERIVATION

In this notes we derive the nonlinear sigma-model (NLσM) action starting from the edge Hamiltonian $\hat{H}$ obtained in Ref. [6]. For simplicity, we perform the derivation at a fixed energy. The action for Grassmann fields $\hat{\psi}$, suitable for constructing the retarded Green function, reads

$$\mathcal{S}_e = -i \int_0^L dy_1 dy_2\, \hat{\bar{\psi}}(y_1) \left( i0 - \hat{H}(y_1, y_2) \right) \hat{\psi}(y_2), \quad \hat{H}(y_1, y_2) = \delta(y_{12}) \left( -iv\partial_{y_2} - vk_0\hat{\sigma}_3 \right) + t^2 \left( \partial_x \Phi \right)^2 \partial^2_{x_1,x_2} \hat{\mathcal{G}}. \tag{S.25}$$

Here $\hat{\psi} = (\psi_\uparrow, -\bar{\psi}_\downarrow)$, $\hat{\sigma}_i$ are Pauli matrices in the Bogoliubov-de Gennes (BdG) space and $\hat{\mathcal{G}}$ is the exact Green's function of a disordered superconductor evaluated at zero energy, $E = 0$, and at the interface $x_{1,2} = z_{1,2} = 0$. It can be represented as [6, 7]

$$\hat{\mathcal{G}}(\boldsymbol{r}_1, \boldsymbol{r}_2; E = 0) = \int \frac{d\varepsilon}{\varepsilon^2 + \Delta^2} \begin{pmatrix} \varepsilon & \Delta \\ \Delta & -\varepsilon \end{pmatrix} \frac{1}{\pi} \operatorname{Im} \mathcal{G}^R_{\mathrm{N}}(\boldsymbol{r}_1, \boldsymbol{r}_2; \varepsilon) = g(\boldsymbol{r}_1, \boldsymbol{r}_2; E = 0)\hat{\sigma}_3 + f(\boldsymbol{r}_1, \boldsymbol{r}_2; E = 0)\hat{\sigma}_1, \tag{S.26}$$

where $g$ and $f$ are the normal and anomalous Green's functions of the disordered superconductor. The key idea is to treat $\hat{\mathcal{G}}$ as an effective random potential, since it depends on the impurity potential $U(\mathbf{r})$ with the following two-point correlation function

$$\langle U(\boldsymbol{r}) U(\boldsymbol{r}') \rangle = \frac{v_{\mathrm{F}}}{2\pi\nu_{\mathrm{F}} l_{\mathrm{sc}}} \delta\left( \boldsymbol{r} - \boldsymbol{r}' \right). \tag{S.27}$$

We then derive an effective action by averaging over disorder. As a first step, it is convenient to gauge out the 2DEG Fermi momentum $k_0$. We introduce transformed fields $\hat{\eta}$

$$\hat{\psi} = e^{i\hat{\sigma}_3 k_0 y} \hat{\eta}, \quad \hat{\bar{\psi}} = \hat{\bar{\eta}} e^{-i\hat{\sigma}_3 k_0 y}. \tag{S.28}$$

After this transformation, Hamiltonian becomes (cf. Eq.(6) from the main text)

$$\hat{H}(y_1, y_2) = \delta(y_{12}) \left( -iv\partial_{y_2} \right) + \hat{V}(y_1, y_2), \tag{S.29}$$

where a random potential

$$\hat{V}(y_1, y_2) = t^2 \left( \partial_x \Phi \right)^2 \partial^2_{x_1,x_2} \Big[ g(\boldsymbol{r}_1, \boldsymbol{r}_2) e^{-i\hat{\sigma}_3 k_0 (y_1 - y_2)} \hat{\tau}_3 + f(\boldsymbol{r}_1, \boldsymbol{r}_2) \cos(k_0(y_1 + y_2)) \hat{\sigma}_1$$
$$+ f(\boldsymbol{r}_1, \boldsymbol{r}_2) \sin(k_0(y_1 + y_2)) \hat{\sigma}_2 \big] \big|_{x_{1,2} = z_{1,2} = 0}\,. \tag{S.30}$$

One can verify that the Hamiltonian satisfies the class C Bogoliubov–de Gennes (charge-conjugation) symmetry, $\hat{H} = -\hat{\sigma}_2 \hat{H}^T \hat{\sigma}_2$. To perform disorder averaging, we employ the replica trick and replicate the action,

$$\mathcal{S}_e = -i \sum_{\alpha=1}^{N_r} \int_0^L dy_1 dy_2\, \hat{\bar{\eta}}_\alpha(y_1) \left( i0 - \hat{H}(y_1, y_2) \right) \hat{\eta}_\alpha(y_2) \tag{S.31}$$

To capture all soft modes, we introduce doubled fields,

$$\hat{\Xi}_\alpha = \frac{1}{\sqrt{2}} \begin{pmatrix} \hat{\eta}_\alpha \\ \hat{\sigma}_2 \hat{\bar{\eta}}^T_\alpha \end{pmatrix}, \quad \hat{\bar{\Xi}}_\alpha = \frac{1}{\sqrt{2}} \left( \hat{\bar{\eta}}_\alpha, \hat{\eta}^T_\alpha \hat{\sigma}_2 \right), \quad \hat{\bar{\Xi}}_\alpha = \hat{\Xi}^T_\alpha \hat{C}, \quad \hat{C} = \begin{pmatrix} 0 & \hat{\sigma}_2 \\ -\hat{\sigma}_2 & 0 \end{pmatrix} = i\hat{s}_2 \hat{\sigma}_2, \tag{S.32}$$

where $\hat{s}_i$ are Pauli matrices in Nambu space. In terms of these fields, the action becomes

$$\mathcal{S}_e = -i \sum_{\alpha=1}^{N_r} \int_0^L dy_1 dy_2\, \hat{\bar{\Xi}}_\alpha(y_1) \left( i0\hat{s}_3 - \hat{H}(y_1, y_2) \hat{s}_0 \right) \hat{\Xi}_\alpha(y_2). \tag{S.33}$$

We now evaluate the disorder-averaged Green's functions. For the normal ($g$) and anomalous ($f$) components, we obtain

$$\langle g(\boldsymbol{r}_1, \boldsymbol{r}_2) \rangle_U = \int \frac{d\varepsilon\ \varepsilon}{\varepsilon^2 + \Delta^2} \frac{1}{\pi} \left\langle \operatorname{Im} \mathcal{G}^R_{\mathrm{N}}(\boldsymbol{r}_1, \boldsymbol{r}_2; \varepsilon) \right\rangle_U = 0, \quad \langle f(\boldsymbol{r}_1, \boldsymbol{r}_2) \rangle_U = -\pi\nu_{\mathrm{F}} \frac{2 \sin p_{\mathrm{F}} |\boldsymbol{r}_1 - \boldsymbol{r}_2|}{p_F |\boldsymbol{r}_1 - \boldsymbol{r}_2|} \tag{S.34}$$

where $p_{\rm F}$ is the Fermi momentum in the superconductor. For $\partial^2_{x_1,x_2} f$ we find

$$\left\langle \partial^2_{x_1,x_2} f \right\rangle_U = -\pi^2 p_{\rm F} \nu_F \delta(y_{12}), \quad \left\langle \hat{V} \right\rangle_U = -\pi^2 t^2 \left(\partial_x \Phi\right)^2 p_{\rm F} \nu_F \left[\cos(2k_0 y_1)\hat{\sigma}_1 + \sin(2k_0 y_1)\hat{\sigma}_2\right] \delta(y_{12}) \tag{S.35}$$

Here, the $\delta$-function should be understood as a sharp peak of width $\sim \lambda_{\rm F} = 1/p_{\rm F}$. We decompose the random potential as $\hat{V} = \langle \hat{V} \rangle + \delta \hat{V}$. Since we are interested in the leading contribution in the limit $p_{\rm F} l_{\rm sc} \gg 1$, it is sufficient to retain only the second cumulant of the disordered Green's function $\mathcal{G}$ [7]. Applying Wick's theorem, we obtain

$$\left\langle \exp\left( i \int dy_1 dy_2 \hat{\bar{\Xi}}_{\alpha'}(y_1) \hat{V}(y_1,y_2) \hat{\Xi}_{\alpha'}(y_2) \right) \right\rangle_U = \exp\left( i \int dy_1 dy_2 \hat{\bar{\Xi}}_{\alpha'}(y_1) \left\langle \hat{V}(y_1,y_2) \right\rangle_U \hat{\Xi}_{\alpha'}(y_2) \right)$$
$$\times \exp\left( -\frac{1}{2} \int \left[ \prod_{i=1}^{4} dy_i \right] \left\langle \hat{\bar{\Xi}}_{\alpha'}(y_1) \delta\hat{V}(y_1,y_2) \hat{\Xi}_{\alpha'}(y_2) \hat{\bar{\Xi}}_{\beta'}(y_3) \delta\hat{V}(y_3,y_4) \hat{\Xi}_{\beta'}(y_4) \right\rangle_U \right), \tag{S.36}$$

where $\alpha'$ and $\beta'$ denote combined replica and Nambu indices. To evaluate the resulting four-fermion term, we use the irreducible disorder averages (see the Supplemental Material of Ref. [6]):

$$\left\langle\!\left\langle \partial^2_{x_1,x_2} f_{12} \cdot \partial^2_{x_3,x_4} f_{34} \right\rangle\!\right\rangle_U = \frac{\pi^2 \nu_{\rm F} p_{\rm F}^2}{2D} \frac{e^{-|y_1-y_2|/\xi}}{|y_1-y_2|} \left(\delta(y_{13})\delta(y_{24}) + \delta(y_{14})\delta(y_{23})\right) = 2 \left\langle\!\left\langle \partial^2_{x_1,x_2} g_{12} \cdot \partial^2_{x_3,x_4} g_{34} \right\rangle\!\right\rangle_U \tag{S.37}$$

for non-diagonal averages we can write $\left\langle\!\left\langle \partial^2_{x_1,x_2} g_{12} \cdot \partial^2_{x_3,x_4} f_{34} \right\rangle\!\right\rangle_U = \left\langle\!\left\langle \partial^2_{x_1,x_2} f_{12} \cdot \partial^2_{x_3,x_4} g_{34} \right\rangle\!\right\rangle_U = 0$. Using the above correlation functions we reproduce the pair correlation function (7) from the main text. In the universal limit $|y_1 - y_2| \gg \xi$, the pair correlation function simplifies to

$$\frac{\pi^2 \nu_{\rm F} p_{\rm F}^2}{2D} \frac{e^{-|y_1-y_2|/\xi}}{|y_1-y_2|} \left(\delta(y_{13})\delta(y_{24}) + \delta(y_{14})\delta(y_{23})\right) \approx \frac{\gamma_{\rm A}}{2} \delta(y_{12}) \left(\delta(y_{13})\delta(y_{24}) + \delta(y_{14})\delta(y_{23})\right), \tag{S.38}$$

where $\gamma_{\rm A} = 2\pi^2 \nu_{\rm F} p_{\rm F}^2 D^{-1} \ln(\xi/l_{\rm sc})$. Using these identities, the four-fermion interaction reduces to

$$-\frac{\gamma_{\rm A}}{4} \int dy \left( \hat{\bar{\Xi}}_{\alpha'} \hat{\sigma}_3 \hat{\Xi}_{\alpha'} \hat{\bar{\Xi}}_{\beta'} \hat{\sigma}_3 \hat{\Xi}_{\beta'} + 2 \hat{\bar{\Xi}}_{\alpha'} \hat{\sigma}_1 \hat{\Xi}_{\alpha'} \hat{\bar{\Xi}}_{\beta'} \hat{\sigma}_1 \hat{\Xi}_{\beta'} \cos^2(2k_0 y) + \hat{\bar{\Xi}}_{\alpha'} \hat{\sigma}_1 \hat{\Xi}_{\alpha'} \hat{\bar{\Xi}}_{\beta'} \hat{\sigma}_2 \hat{\Xi}_{\beta'} \sin(4k_0 y) + \right.$$
$$\left. \hat{\bar{\Xi}}_{\alpha'} \hat{\sigma}_2 \hat{\Xi}_{\alpha'} \hat{\bar{\Xi}}_{\beta'} \hat{\sigma}_1 \hat{\Xi}_{\beta'} \sin(4k_0 y) + 2 \hat{\bar{\Xi}}_{\alpha'} \hat{\sigma}_2 \hat{\Xi}_{\alpha'} \hat{\bar{\Xi}}_{\beta'} \hat{\sigma}_2 \hat{\Xi}_{\beta'} \sin^2(2k_0 y) \right). \tag{S.39}$$

In the universal regime, rapidly oscillating contributions can be neglected, as well as the average potential $\langle \hat{V} \rangle$. This yields

$$\overline{\mathcal{S}}_e^{\rm dis} = -\frac{\gamma_{\rm A}}{4} \int dy \left( 2 \hat{\bar{\Xi}}^n_{\alpha'} \hat{\Xi}^m_{\alpha'} \hat{\bar{\Xi}}^m_{\beta'} \hat{\Xi}^n_{\beta'} - \hat{\bar{\Xi}}^n_{\alpha'} \hat{\Xi}^n_{\alpha'} \hat{\bar{\Xi}}^m_{\beta'} \hat{\Xi}^m_{\beta'} \right) \tag{S.40}$$

where $n, m$ are indices in BdG space. The second term vanishes due to the definition of the $\Xi$ fields. Next, proceeding along standard lines (see Ref. [8]), we arrive at the nonlinear sigma-model action

$$\mathcal{S}[Q] = \frac{1}{2} \int_0^L dy \, \mathrm{Tr}\, \Lambda T \partial_y T^{-1} - \frac{\ell_A}{8} \int_0^L dy \, \mathrm{Tr}(\partial_y Q)^2. \tag{S.41}$$

where the matrix field $Q = T^{-1} \hat{s}_3 T$ satisfies the standard NL$\sigma$M constraint $Q^2 = 1$ as well as the charge-conjugation symmetry $Q = -\hat{s}_2 Q^T \hat{s}_2$. Consequently, $Q$ belongs to the target manifold of symmetry class C [9]. We emphasize that the $Q$ matrix defined here is related to that used in the main text and Ref. [8] by a non-unitary rotation (see Ref. [10]).

The extension of the above derivation to the Matsubara frequency formalism is straightforward and reproduces the NL$\sigma$M action given by Eq. (8) in the main text.

## S.IV. TRANSFER MATRIX APPROACH FOR $\mathbf{T}_{ij}^{\alpha\beta}$

In this section, we derive the transmission amplitudes using the transfer-matrix approach. A key simplification arises from the chiral nature of the edge states, which allows us to consider only forward propagation without backscattering.

We begin by introducing the propagation matrices. We work in the standard BdG spinfull basis, where the vector of the amplitudes has the form $(a_e^{\uparrow}, a_e^{\downarrow}, a_h^{\uparrow}, a_h^{\downarrow})$. For free ballistic motion along the edge of the 2DEG [see Fig.2 in the main text], the evolution is described by diagonal phase matrices. Specifically, for segment $A$ (reservoir 1 to right constriction), segment $B$ (right constriction to superconductor), segment $C$ (superconductor to left constriction), and segment $D$ (left constriction to reservoir 4), we have, e.g., $\hat{A} = \mathrm{diag}\{e^{i\varphi_\uparrow^{(A)}}, e^{i\varphi_\downarrow^{(A)}}, e^{-i\varphi_\uparrow^{(A)}}, e^{-i\varphi_\downarrow^{(A)}}\}$, with analogous expressions for the other segments. At the constrictions, scattering is described by reflection and transmission matrices

$$\hat{R}_i = \begin{pmatrix} \hat{R}_i^{(e)} & 0 \\ 0 & \hat{R}_i^{(h)} \end{pmatrix}, \quad \hat{T}_i = \begin{pmatrix} \hat{T}_i^{(e)} & 0 \\ 0 & \hat{T}_i^{(h)} \end{pmatrix} \tag{S.42}$$

where $i$ labels the QPC. The electron and hole blocks are related by particle–hole symmetry:

$$\hat{R}_i^{(e)} = \left(\hat{R}_i^{(h)}\right)^* = \begin{pmatrix} r_i^{\uparrow\uparrow} & 0 \\ 0 & r_i^{\downarrow\downarrow} \end{pmatrix} \tag{S.43}$$

and similarly for $\hat{T}_i^{(e)}$ with $r \to t$. For inner QPCs, we use a prime to distinguish the corresponding matrices. In the absence of spin-flip processes, all off-diagonal elements in spin space vanish, and the matrices $\hat{R}_i^{(e,h)}$ and $\hat{T}_i^{(e,h)}$ are diagonal. We assume that the QPCs act as ideal spin filters, such that $r_i^{\uparrow\uparrow} = 1$, $r_i^{\downarrow\downarrow} = 0$ and $t_i^{\uparrow\uparrow} = 0$, $t_i^{\downarrow\downarrow} = 1$. For the inner QPCs, the spin indices are reversed. Finally, propagation along the S–2DEG interface is described by the scattering matrix $\hat{S}$:

$$\hat{S} = \begin{pmatrix} \hat{S}^{(ee)} & \hat{S}^{(eh)} \\ \hat{S}^{(eh)} & \hat{S}^{(hh)} \end{pmatrix}, \quad \hat{S}^{(ee)} = \begin{pmatrix} t_{ee}^{\uparrow\uparrow} & 0 \\ 0 & t_{ee}^{\downarrow\downarrow} \end{pmatrix}, \quad \hat{S}^{(eh)} = \begin{pmatrix} 0 & t_{eh}^{\uparrow\downarrow} \\ t_{eh}^{\downarrow\uparrow} & 0 \end{pmatrix} \tag{S.44}$$

This structure follows from spin conservation at the 2DEG-S interface. Although our results are insensitive to the specific form of the transmission amplitudes $t$, for completeness we note that calculations of the $\hat{S}$-matrix elements in different regimes can be found in the literature: in the diffusive regime without a Zeeman field in Ref. [6], in the ballistic regime without a Zeeman field in Ref. [11], in the ballistic regime with a Zeeman field in Ref. [1], and for related ballistic calculations at ferromagnet-S interfaces in Refs. [12–14]. We notice that the transmission coefficients in Eq. (9) in the main text decompose into spin components as $\mathsf{T}_{ij}^{\alpha\beta} = \sum_{\mathsf{s},\mathsf{s}'=\uparrow,\downarrow} {}^{\mathsf{s}\mathsf{s}'}\mathsf{T}_{ij}^{\alpha\beta}$, where ${}^{\mathsf{s}\mathsf{s}'}\mathsf{T}_{ij}^{\alpha\beta}$ describes scattering from spin $\mathsf{s}'$ to spin $\mathsf{s}$.

As an example, consider the transmission from terminal 1 to terminal 3. The corresponding transfer matrix is given by

$$\hat{T}_{31} = \hat{T}_2 \hat{C} \hat{S} \hat{B} \hat{R}_1 \hat{A} = \begin{pmatrix} 0 & 0 & 0 & 0 \\ 0 & 0 & e^{-i(\varphi_\uparrow^A + \varphi_\uparrow^B - \varphi_\downarrow^C)} t_{eh}^{\downarrow\uparrow} & 0 \\ 0 & 0 & 0 & 0 \\ e^{i(\varphi_\uparrow^A + \varphi_\uparrow^B - \varphi_\downarrow^C)} t_{he}^{\downarrow\uparrow} & 0 & 0 & 0 \end{pmatrix}. \tag{S.45}$$

Thus, the only nonvanishing transmission probabilities are ${}^{\downarrow\uparrow}\mathsf{T}_{31}^{he} = \left|t_{he}^{\downarrow\uparrow}\right|^2$ and ${}^{\downarrow\uparrow}\mathsf{T}_{31}^{eh} = \left|t_{eh}^{\downarrow\uparrow}\right|^2$. We note that transport from terminal 1 to terminal 3 occurs exclusively via Andreev processes. The probabilities for other transport processes can be found similarly. The non-zero ones are given by

$${}^{\uparrow\uparrow}\mathsf{T}_{41}^{\alpha\alpha} = \left|t_{\alpha\alpha}^{\uparrow\uparrow}\right|^2, \quad {}^{\downarrow\downarrow}\mathsf{T}_{32}^{\alpha\alpha} = \left|t_{\alpha\alpha}^{\downarrow\downarrow}\right|^2, \quad {}^{\uparrow\downarrow}\mathsf{T}_{42}^{\bar{\alpha}\alpha} = \left|t_{\bar{\alpha}\alpha}^{\uparrow\downarrow}\right|^2. \tag{S.46}$$

Using Eq. (9) of the main text, we evaluate the currents in terminals 3 and 4. Assuming $\mu_3 = \mu_4 = 0$ and working in the linear-response regime, we obtain

$$I_3 = -\frac{e}{2h}\left[\left(-\left|t_{eh}^{\downarrow\uparrow}\right|^2 - \left|t_{he}^{\downarrow\uparrow}\right|^2\right)\mu_1 + \left(\left|t_{ee}^{\downarrow\downarrow}\right|^2 + \left|t_{hh}^{\downarrow\downarrow}\right|^2\right)\mu_2\right], \tag{S.47}$$

$$I_4 = -\frac{e}{2h}\left[\left(\left|t_{ee}^{\uparrow\uparrow}\right|^2 + \left|t_{hh}^{\uparrow\uparrow}\right|^2\right)\mu_1 + \left(-\left|t_{eh}^{\uparrow\downarrow}\right|^2 - \left|t_{he}^{\uparrow\downarrow}\right|^2\right)\mu_2\right]. \tag{S.48}$$

where all amplitudes are evaluated at zero energy, $E = 0$. In the low-bias regime, where no quasiparticle current flows into the superconductor, probability conservation implies $|t^{\mathsf{ss}}_{ee}|^2 + \left|t^{\mathsf{s}'\mathsf{s}}_{he}\right|^2 = 1$ with an analogous relation for hole processes. Using these identities and defining $|t^{\mathsf{ss}}_{ee}|^2 = |t^{\mathsf{ss}}_{hh}|^2 = \mathsf{T}_{\mathrm{N}} = 1 - \mathsf{T}_{\mathrm{A}}$ one recovers Eq. (10) of the main text.




\end{document}